\documentclass[prletter,showpacs,twocolumn,typeset,superscriptaddress]{revtex4}
\usepackage{amsfonts}
\usepackage{amsmath}
\usepackage{amssymb}
\usepackage{graphicx}
\usepackage{epsfig}
\usepackage{citesort}

\usepackage{dcolumn}
\usepackage{bm}

\begin{document}

\preprint{APS}

\title{Decoherence assisting a measurement-driven quantum evolution process}

\author{L. Roa}
\email{lroa@udec.cl}
\affiliation{Center for Quantum Optics and Quantum Information, Departamento de F\'{\i}sica,
Universidad de Concepci\'{o}n, Casilla 160-C, Concepci\'{o}n,
Chile.}
\author{G. Olivares}
\affiliation{Center for Quantum Optics and Quantum Information, Departamento de F\'{\i}sica,
Universidad de Concepci\'{o}n, Casilla 160-C, Concepci\'{o}n,
Chile.}
\date{\today}

\begin{abstract}
We study the problem of driving an unknown initial mixed quantum
state onto a known pure state without using unitary
transformations. This can be achieved, in an efficient manner, with the help of sequential
measurements on at least two unbiased bases. However here
we found that, when the system is affected by a decoherence
mechanism, only one observable is required in order to achieve
the same goal. In this way the decoherence can assist the process.
We show that, depending on the sort of decoherence, the process can
converge faster or slower than the method implemented by means
of two complementary observables.
\end{abstract}

\maketitle

During the last two decades a major research effort has been
conducted in the emerging field of quantum information theory
\cite{Nielsen}. Much of this activity started with the observation
that the capacity of physical systems to process, store and
transmit information depends on their classical or quantum nature
\cite{Landauer}. Algorithms based on
the laws of Quantum Mechanics show an enhancement of information
processing capabilities over their classical counterparts. A large
collection of quantum communication protocols such as quantum
teleportation \cite{Bennett}, entanglement swapping
\cite{Zukowsky}, quantum cloning \cite{Wootters,Duan} and quantum
erasing \cite{Pati} reveal new forms of transmitting and storing
classical and quantum information. Most of these protocols have
already been experimentally implemented
\cite{Zhao1,Zhao2,Pan1,Pan2,Bouwmeester1,Boschi,Bouwmeester2}.

A common assumption concerning quantum algorithms and quantum
communication protocols is the capacity of performing transformations
belonging to a fixed but arbitrary set of unitary transformations
together with measurements on a given basis. An interesting application
in this context is \textit{quantum information dilution} \cite{Ziman}.
Here, an arbitrary unknown state of a two-dimensional quantum system is
asymptotically driven onto a particular state by interacting with a
finite reservoir of two-dimensional quantum systems. This is implemented
by means of a sequence of unitary swapping interactions.
Another application is the \textit{measurement-driven
quantum evolution} process \cite{Roa}, which is succinctly described,
in a two-dimensional Hilbert space, as follows:
Suppose that initially the
system is in an unknown state $\rho_0$. The goal consists in
mapping this state onto the known pure target state
$|\varsigma\rangle$.

In order to accomplish this task we require a non-degenerate
observable $\hat{\varsigma}$ with eigenstates
$\{|\varsigma\rangle,|\varsigma_\perp\rangle\}$. So, the target
state must belong to the spectral decomposition of $\hat{\varsigma}$.

A measurement of the $\hat{\varsigma}$ observable when the system is in the $\rho_0$
state projects the system onto the target state $|\varsigma\rangle$
with probability $p=\langle\varsigma|\rho_0|\varsigma\rangle$. In this
case the process succeeds and no further action is required.
However, the process fails with probability $1-p$ when the measurement projects
the system onto the $|\varsigma_{\perp}\rangle$ state. Since this
state cannot be projected onto $|\varsigma\rangle$ by measuring
$\hat{\varsigma}$, it will be necessary to introduce a second
observable $\hat{\theta}$ whose non-degenerate eigenstates are
denoted by $|0\rangle$ and $|1\rangle$.

Failing the first measurement of $\hat\varsigma$, a measurement
of $\hat{\theta}$ projects the $|\varsigma_{\perp}\rangle$ state
onto either the state $|0\rangle$ or  $|1\rangle$. Since, in principle, both states have
a component on the $|\varsigma\rangle$ state, a second measurement
of $\hat{\varsigma}$ allows again to project, with a certain
probability, onto the target state $|\varsigma\rangle$.
So, the success probability $p_{s}$ of mapping the initial
state $\rho_0$ onto $|\varsigma\rangle$, the target state, after
applying the consecutive measurement processes
$[M(\hat{\varsigma})M(\hat{\theta})]^NM(\hat{\varsigma})$, that is, a
measurement of $\hat{\varsigma}$ followed by $N$ measurement
processes each one composed of $\hat{\theta}$ followed by
$\hat{\varsigma}$, is
\begin{equation}
p_{s,N}  = 1-\langle\varsigma_{\perp}|\rho_0|\varsigma_{\perp}\rangle\left(
1-2| \langle 0|\varsigma\rangle|^{2}| \langle\varsigma|1\rangle| ^{2}\right)^{N}.   \label{pnotheta}
\end{equation}
This expression
(\ref{pnotheta}) indicates that the success probability $p_{s,N}$
can be maximized by choosing $| \langle 0|\varsigma\rangle|=|
\langle\varsigma|1\rangle|=1/\sqrt{2}$ that corresponds to the definition of mutually unbiased bases (MUB).
It was observed \cite{Roa} that, for MUB, $p_{s,\max}$ quickly converges to $1$ almost
independently of $\langle\varsigma_\perp|\rho_0|\varsigma_\perp\rangle$ even if the
initial state $\rho_0$ belongs to a subspace orthogonal to
$|\varsigma\rangle$.

In this letter we suppose the constraint that no other observable
different from $\hat\varsigma$ can be implemented in the process.
In this case the decoherence can act as an indirect mechanism
which allows taking the state partially out from the
$|\varsigma_\perp\rangle$ direction. Therefore, a first
measurement of the $\hat{\varsigma}$ observable onto the $\rho_0$
state projects the state of the system onto the target state $|\varsigma\rangle$
with probability $p_1=\langle\varsigma|\rho_0|\varsigma\rangle$.
However, the process fails with probability $1-p_1$ when the
measurement projects the system onto the undesired state
$|\varsigma_{\perp}\rangle$;  then a second measurement of
$\hat\varsigma$ is done again after a time $t$ from the first one.
Due to the decoherence mechamism, the $|\varsigma_{\perp}\rangle$ state
evolves to $\rho_t$ and so the probability of projecting it onto the
target state is $p_2=\langle\varsigma|\rho_t|\varsigma\rangle$.
The probability $p^{\prime}$ that this procedure fails in a first
measurement of $\hat{\varsigma}$ but is successful after a second
one is $p^\prime=(1-p_1)p_2$. Let $M(\hat{\varsigma})$ denote a
measurement of the $\hat{\varsigma}$ observable; then the success
probability in the sequence of measurements
$M(\hat{\varsigma})e^{Lt}M(\hat{\varsigma})$ is given by
\begin{equation}
p_{s,2}=p_1+p^\prime=\langle\varsigma|\rho_0|\varsigma\rangle+
(1-\langle\varsigma|\rho_0|\varsigma\rangle)\langle\varsigma|\rho_t|\varsigma\rangle,
\end{equation}
where $e^{Lt}$ is the generic notation of the unavoidable
decoherence mechanism suffered by the two-level system in the state
$|\varsigma_\perp\rangle$ between two consecutive measurement
processes separated by a time $t$, in such a way that
\begin{equation}
\rho_t=e^{Lt}(|\varsigma_\perp\rangle\langle\varsigma_\perp |).  \nonumber
\end{equation}

Similarly, the success probability $p_{s,N}$ of mapping the initial
state $\rho_0$ onto $|\varsigma\rangle$ after applying $N$
consecutive measurement processes of $\hat{\varsigma}$,
all of them separated by a time $t$, is given by
\begin{equation}
p_{s,N}=1-\langle\varsigma_\perp|\rho_0|\varsigma_\perp\rangle
\langle\varsigma_\perp|\rho_t|\varsigma_\perp\rangle^{N-1}.  \label{d}
\end{equation}
In principle the diagonal element satisfies the inequality $0\le
\langle\varsigma_\perp|\rho_t|\varsigma_\perp\rangle\le 1$ at all
time.

\begin{figure} [t]
\includegraphics[angle=360,width=0.40\textwidth]{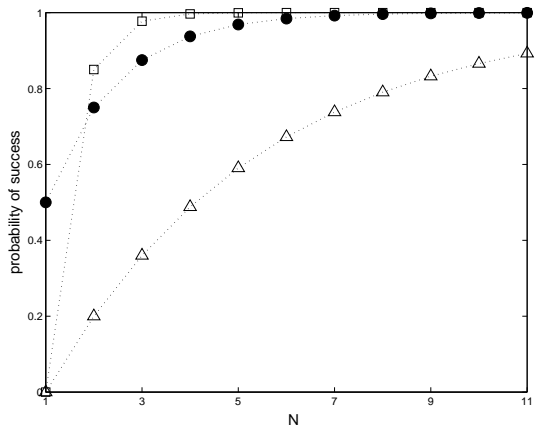}
\caption{Probability of success as a function of $N$ with
$\langle\varsigma_\perp|\rho_0|\varsigma_\perp\rangle=1$, when two
unbiased bases are applied (full circle), and when the decoherence
mechanism is considered with
$\langle\varsigma_\perp|\rho_t|\varsigma_\perp\rangle=0.15$ (square)
and with $\langle\varsigma_\perp|\rho_t|\varsigma_\perp\rangle=0.8$
(triangle).} \label{figure1}
\end{figure}

Figure \ref{figure1} shows the probability of success as a
function of $N$ when two unbiased bases are applied (full
circle), Eq. (\ref{pnotheta}), and when the decoherence mechanism is
considered, Eq. (\ref{d}). En particular we have considered two
values of the diagonal element:
$\langle\varsigma_\perp|\rho_t|\varsigma_\perp\rangle=0.15$
(square) and
$\langle\varsigma_\perp|\rho_t|\varsigma_\perp\rangle=0.8$
(triangle).

From Figure \ref{figure1} we see that, depending on the value of
$\langle\varsigma_\perp|\rho_t|\varsigma_\perp\rangle$, the process
with decoherence mechanism can converge either faster (square) or
slower (triangle) than the procedure by means of unbiased bases (full circle).

We can deduce from Eqs. (\ref{pnotheta}) and (\ref{d}) that, if
\begin{equation}
\langle\varsigma_\perp|\rho_t|\varsigma_\perp\rangle \leq
2^{-N/(N-1)},   \label{swd}
\end{equation}
then the process with decoherence mechanism (DM) converges faster
than the procedure realized by means of mutually unbiased bases
(MUB), after $N$ measurement procedures. In other words,
in order for the DM procedure to converge faster than the MUB process,
it is required that at least
$\langle\varsigma_\perp|\rho_t|\varsigma_\perp\rangle < 1/2$. The
DM process will be fastest when
$\langle\varsigma_\perp|\rho_t|\varsigma_\perp\rangle < 1/4$ which
corresponds to $N=2$.

For instance, let us consider a two-level system subjected to pure
dephasing decoherence. This effective mechanism \cite{Bose1} is
caused by an interaction with at least one boson mode \cite{Zurek} described
by the coupling hamiltonian $g(b+b^\dagger )\sigma_z$, where
$\sigma_z$ is the \textit{z}-component of the $\mathbf{\sigma}$
spin-1/2 operator with eigenstates $|0\rangle$ and $|1\rangle$.
$b$ and $b^\dagger $ are the boson annihilation and creation
operators respectively, and $g$ gives account of the spin-boson
effective coupling strength.

Considering the boson mode to be initially
in a thermal state, the
$\langle\varsigma_\perp|\rho_t|\varsigma_\perp\rangle$ diagonal
element of the $\rho_t$ reduced spin density operator is given by
\begin{equation}
\langle\varsigma_\perp|\rho_t|\varsigma_\perp\rangle=1-2\left\{1-A_T(t)\right\}
|\langle 0|\varsigma_\perp\rangle|^2|\langle 1|\varsigma_\perp\rangle|^2, \label{ele}
\end{equation}
with
\begin{equation}
A_T(t)=\frac{e^{-2(gt)^2}}{1+\langle
n\rangle_T}\sum_{n=0}^\infty\left(\frac{\langle
n\rangle_T}{1+\langle n\rangle_T}\right)^nL_n(4g^2t^2), \nonumber
\end{equation}
where $L_n$ is the Laguerre polynomial, $\langle
n\rangle_T=(e^{\hbar\omega/k_BT}-1)^{-1}$ is the mean thermal
boson number, $k_B$ is the Boltzmann constant, $T$ is the
absolute temperature, $\omega$ is the frequency of the boson
mode, and $\hbar$ is the universal Planck constant. In the low
temperature regimen ($T\sim 0$) the (\ref{ele}) element goes to
$1-2(1-e^{-2(gt)^2})|\langle 0|\varsigma_\perp\rangle|^2|\langle
1|\varsigma_\perp\rangle|^2$ which is always higher than or equal to
$1/2$. In the high temperature regime ($\langle n\rangle_T\gg 1$)
the (\ref{ele}) element goes to $1-2(1-\delta_{t,0})(|\langle
0|\varsigma_\perp\rangle|^2|\langle 1|\varsigma_\perp\rangle|^2$
which also is always higher than or equal to $1/2$. It can be easily
shown that the density operator of a two-level system under the
effect of any pure dephasing mechanism satisfies
$\langle\varsigma_\perp|\rho_t|\varsigma_\perp\rangle\leq1/2$ at
all times $t$ and for any $|\varsigma_\perp\rangle$ initial state.

As a second example let us consider a two-level system interacting
resonantly with a quantized boson mode through the Jaynes-Cummings
model \cite{Jaynes},
$g(b\sigma_++b^\dagger\sigma_-)$. Similarly to
the previous example, the field is initially in a thermal state at
$T$ temperature \cite{Bose}. In this case the
$\langle\varsigma_\perp|\rho_t|\varsigma_\perp\rangle$ diagonal
element becomes
\begin{widetext}
\begin{eqnarray}
\langle \varsigma _{\perp }|\rho _{t}|\varsigma _{\perp }\rangle
&=&\frac{1}{1+\langle n\rangle_T}\sum_{n=0}^{\infty }\left\{ \left[ \left\vert \langle 0|\varsigma _{\perp
}\rangle \right\vert ^{2}\cos \left( \sqrt{n}gt\right) +\left\vert \langle
1|\varsigma _{\perp }\rangle \right\vert ^{2}\cos \left( \sqrt{n+1}gt\right)
\right] ^{2}\right.   \nonumber \\
&&+\left. \left\vert \langle \varsigma _{\perp }|0\rangle \right\vert
^{2}\left\vert \langle 1|\varsigma _{\perp }\rangle \right\vert ^{2}\left[
\sin ^{2}\left( \sqrt{n}gt\right) +\sin ^{2}\left( \sqrt{n+1}gt\right)
\right] \right\} \left(\frac{\langle n\rangle_T}{1+\langle n\rangle_T}\right)^n. \label{ce}
\end{eqnarray}
\end{widetext}

\begin{figure} [t]
\includegraphics[angle=360,width=0.40\textwidth]{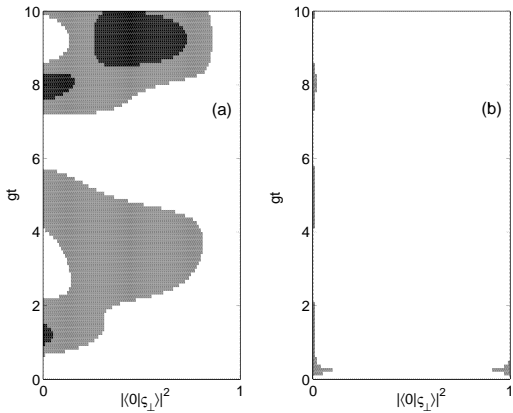}
\caption{Diagonal element,
$\langle\varsigma_\perp|\rho_{t}|\varsigma_\perp\rangle$, as a
function of $gt$ and $|\langle 0|\varsigma_\perp\rangle|^2$ with:
(a) $\langle n\rangle_T=1$ and (b) $\langle n\rangle_T=1000$.
White means that
$\langle\varsigma_\perp|\rho_t|\varsigma_\perp\rangle \geq 1/2$,
grey means that
$1/2>\langle\varsigma_\perp|\rho_t|\varsigma_\perp\rangle > 1/4$,
and black means that $1/4\geq
\langle\varsigma_\perp|\rho_t|\varsigma_\perp\rangle \geq 0.06$.}
\label{figure2}
\end{figure}

Figure \ref{figure2} shows the dynamics of the (\ref{ce}) diagonal
element for all $|\langle 0|\varsigma_\perp\rangle|^2$ values. We
have considered: (a) low temperature with $\langle n\rangle_T=1$
and (b) high temperature with $\langle n\rangle_T=100$. Clearly
there are some zones of $(|\langle
0|\varsigma_\perp\rangle|^2,gt)$ for which the diagonal (\ref{ce})
element is under the $1/4$ (black zones) which means that at $N=2$
the decoherence measurement-driven process starts to converge
faster than the unbiased bases process. This favorable behavior
for the DM procedure is widely and strongly
($1/4>\langle\varsigma_\perp|\rho_t|\varsigma_\perp\rangle
> 0$) present at low temperature whereas for high temperature
there are zones where the a weak effect,
($1/2>\langle\varsigma_\perp|\rho_t|\varsigma_\perp\rangle >
1/4$), is present.

\begin{figure} [t]
\includegraphics[angle=360,width=0.40\textwidth]{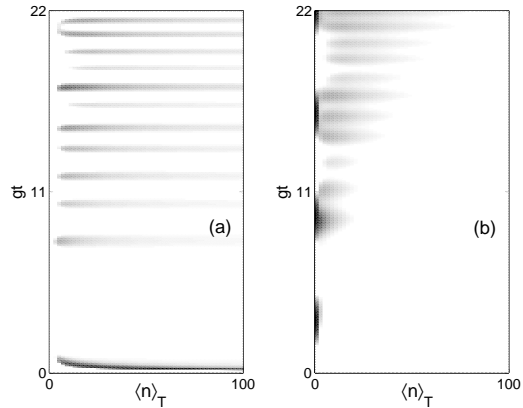}
\caption{Diagonal element,
$\langle\varsigma_\perp|\rho_{t}|\varsigma_\perp\rangle$, as a
function of $gt$ and $\langle n\rangle_T$ with: (a) $|\langle
0|\varsigma_\perp\rangle|^2=1$ and (b) $|\langle
0|\varsigma_\perp\rangle|^2=1/2$. White stands for
$\langle\varsigma_\perp|\rho_t|\varsigma_\perp\rangle \geq 1/2$
whereas the gray degradation goes from $1/2$ (white) up to $\sim 0.18$ (black) for graphic (a) and
up to $\sim 8\times 10^{-6}$ (black) for graphic (b).} \label{figure3}
\end{figure}

Figure \ref{figure3} shows the
$\langle\varsigma_\perp|\rho_{t}|\varsigma_\perp\rangle$ diagonal
element as a function of $gt$ and $\langle n\rangle_T$, with: (a)
$|\langle 0|\varsigma_\perp\rangle|^2=0$, and (b) $|\langle
0|\varsigma_\perp\rangle|^2=0.5$. For $|\langle 0|\varsigma_\perp\rangle|^2=0$,
the more efficient behavior zones, this is
$\langle\varsigma_\perp|\rho_{t}|\varsigma_\perp\rangle\leq1/4$,
appear and it reamains only for
small $gt$ at high $\langle n\rangle_T$ values; meanwhile, for $|\langle
0|\varsigma_\perp\rangle|^2=0.5$ a efficient behavior zones appear and
it desapear at high $\langle n\rangle_T$ values.

In summary, we have compared two procedures which allow driving
an unknown quantum state toward a known pure state by means of von
Neumann measurements procedures only. The first process is based on
a sequence of measurements of two non-commuting observables. The
success probability turns out to be most efficient under the
condition that the observables define mutually unbiased bases. The
second process which we present here makes use of a decoherence
mechanism instead of a second observable. The decoherence
mechanism allows taking the state out from the direction orthogonal
to that of the target state. Which of these processes is optimum
depends on the
$\langle\varsigma_\perp|\rho_t|\varsigma_\perp\rangle$ value with
respect to $2^{-N/(N-1)}$. When this is smaller than o equal to
$2^{-N/(N-1)}$, the second procedure converges to $1$ faster than
the first one after the $N$ measurements. We have shown that the pure
dephasing mechanism can not be more efficient than the procedure
which makes use of two mutually unbiased bases; whereas a
non-dispersive decoherence mechanism described by a
Jaynes-Cummings model can be more efficient than the procedure
which makes use of two mutually unbiased bases. Thus the
decoherence can assist the process and, for a non-dispersive
decoherence regimen, it can accelerate the convergence of the
process towards the success. It is worth noting that there are
some states more stable under the decoherence mechanism than others
\cite{AK}. Specifically it was found \cite{AK} that if one
state, for instance the target, has a certain decoherence time scale,
its orthogonal direction can have either a higher or a smaller one,
depending on the reservoir parameters.

Further studies could involve other natural interactions. For example,
one could study a more real model for decoherence mechanism, this is,
considering an infinite
set of mode near or far from the resonance. It could be also generalized
considering an $d$-dimensional Hilbert space. 

\begin{acknowledgments}
This work was supported by Grants Milenio ICM P02-49F and FONDECyT No. 1030671.
The authors thank Carlos Saavedra, Aldo Delgado and Loreto Ladr\'{o}n de Guevara for valuble dicussions. 
\end{acknowledgments}

\end{document}